\documentclass[preprint,aps,showpacs,preprintnumbers]{revtex4}%
\usepackage{bm}
\usepackage{amsmath}
\usepackage{amsfonts}
\usepackage{graphicx,color}
\usepackage{color}%
\usepackage{amssymb}%
\usepackage{graphicx}
\providecommand{\U}[1]{\protect\rule{.1in}{.1in}}
\DeclareMathAlphabet\mathbfcal{OMS}{cmsy}{b}{n}
\begin{document}
\title{Pseudo-$\mathcal{PT}$ \ symmetric Dirac equation : effect of a new mean spin
angular momentum operator on Gilbert damping}
\author{Y. Bouguerra, S. Mehani, K.\ Bechane and M. Maamache }
\affiliation{Laboratoire de Physique Quantique et Syst\`{e}mes Dynamiques, Facult\'{e} des
Sciences, Universit\'{e} Ferhat Abbas S\'{e}tif 1, S\'{e}tif 19000, Algeria}
\author{P. -A. Hervieux}
\affiliation{Institut de Physique et Chimie des Mat{\'e}riaux de Strasbourg, CNRS and
Universit{\'e} de Strasbourg BP 43, F-67034 Strasbourg, France}
\date{\today }

\begin{abstract}
The pseudo-$\mathcal{PT}$ \ symmetric Dirac equation is proposed and analyzed
by using a non-unitary Foldy-Wouthuysen transformations. A new spin operator
$\mathcal{PT}$ \textbf{\ }symmetric expectation value (called the mean spin
operator) for an electron interacting with a time-dependent electromagnetic
field is obtained. We show that spin magnetization - which is the quantity
usually measured experimentally - is not described by the standard spin
operator but by this new mean spin operator to properly describe magnetization
dynamics in ferromagnetic materials and the corresponding equation of motion
is compatible with the phenomenological model of the Landau-Lifshitz-Gilbert
equation (LLG).

\end{abstract}

\pacs{Pseudo PT Symmetry; Non-Hermitian Dirac equation; Foldy-Wouthuysen
transformation; Landau-Lifshitz-Gilbert equation}
\maketitle

In the field of micromagnetism${\normalsize ,}$ which provides the physical
framework for understanding and simulating ferromagnetic
materials${\normalsize ,}$ there is a fundamental unsolved problem which is
the microscopic origin of the intrinsic Gilbert damping. However, this damping
mechanism has been introduced phenomenologically by T. L. Gilbert in 1955 for
describing the spatial and temporal evolution of the magnetization (known as
the LLG equation), a vector field which determines the properties of
ferromagnetic materials on the sub-micron length scale \cite{Gil55}. Let us
stress that this equation leads to the conservation of the magnetization
modulus. This phenomenological model has since been validated by numerous
experimental data and constitutes the foundation of micromagnetism
\cite{Kro03}. Moreover, magnetic damping plays a crucial role in the operation
of magnetic devices\textbf{.\ }The scattering theory can be used to compute
the Gilbert damping tensor \cite{Brataas}.

Spin is a quantum concept \cite{Jean} that arises naturally from the Dirac
theory and is associated with the operator $\hat{\mathbf{\Sigma}}^{\mathrm{D}%
}\equiv\frac{1}{2i}\mathbf{\alpha}\wedge\mathbf{\alpha}=\left(
\begin{array}
[c]{cc}%
\mathbf{\sigma} & 0\\
0 & \mathbf{\sigma}%
\end{array}
\right)  $ where $\mathbf{\alpha}=\left(
\begin{array}
[c]{cc}%
0 & \mathbf{\sigma}\\
\mathbf{\sigma} & 0
\end{array}
\right)  $ and $\mathbf{\sigma}$ are the usual $2\times2$ Pauli matrices
\cite{Sak67}. \ \textbf{For a classical version of the spin (see Supplementary
Materials). }Usually, spin magnetization $\mathbf{M}(\mathbf{r},t)$ (the
quantity which is experimentally measured) is defined as the expectation value
of the spin angular momentum given by $\mu_{\mathrm{B}}\left\langle
\Psi^{\mathrm{D}}|\hat{\mathbf{\Sigma}}^{\mathrm{D}}|\Psi^{\mathrm{D}%
}\right\rangle $ with $\mu_{\mathrm{B}}\equiv\frac{e\hbar}{2m}$ the Bohr's
magneton $(e<0)$ and where $\Psi^{\mathrm{D}}$ is{\normalsize \ \ }%
$a${\normalsize \ }solution of the Dirac equation. Indeed, in most magnetic
materials the orbital moment is quenched and therefore magnetism is only due
to the spins \cite{Blu01}.

While for a free electron the spin angular momentum in the Heisenberg picture
is not a constant of motion $\left(  \dfrac{d\mathbf{\hat{\Sigma}}%
^{\mathrm{D}}}{dt}\neq0\right)  $ \cite{Sak67}, there exists another spin
operator $\hat{\overline{\Sigma}}^{\mathrm{D}}$\textbf{, }%
considered\ to\ be\ a\ constant\ of\ motion\textbf{,} (called the mean spin
operator \cite{Fol50}) \textbf{\ }$\left(  \dfrac{d\hat{\overline
{\mathbf{\Sigma}}}^{\mathrm{D}}}{dt}=0\right)  $. In the presence of an
electromagnetic field, which is relevant for exploring the microscopic origin
of the Landau-Lifshitz-Gilbert (LLG) equation,\textbf{\ }a satisfactory result
has not yet given.

Knowing that the spinors in the Dirac theory consist of four components, it is
important to check whether the Dirac equation yields physically reasonable
results in the non-relativistic\ expansion case and to show that the Dirac
equation reproduces the two-component Pauli equation. We transform the
Hamiltonian in such a way that all operators of the type $\alpha$\ that couple
the large to the small components will be removed. This can be achieved by a
Foldy Wouthuysen transformation \cite{Fol50,Gre00,bjorken} which is a
non-relativistic\ expansion of the Hamiltonian
into\ series\ of\ the\ particle$\prime$s\ Compton\ wave\ lengths\ $\lambda
_{\mathrm{C}}\equiv\frac{h}{mc}$.

Hickey and Moodera \cite{Hic09} have proposed that the spin-orbit interaction,
which arises from the non-relativistic expansion of the Dirac equation, may be
responsible for the intrinsic ferromagnetic line\textrm{\ }\textit{width}. In
their work, the term containing the curl of the electric field when coupled to
Maxwell's equations lead to a time-varying magnetic induction${\normalsize ,}$
and the theoretical methods employed involve previously developed formalisms
in which an effective non-Hermitian and time-dependent Hamiltonian is used.
However, the non-Hermiticity of the Hamiltonian imposes new rules which are
modified with respect to those of standard quantum mechanics. This fact was
not explicitly taken into account by the authors of \cite{Hic09} and
therefore, their derivation of the intrinsic damping process is unfortunately
incorrect. Moreover, there is another fundamental issue which emerges from
this work \cite{Hic09} concerning how to properly perform the coupling between
the classical Maxwell equations and the quantum evolution resulting from the
non-relativistic limit of the Dirac equation. In what\ follows, we show how to
overcome this difficulty by using the well-known correspondence principle.
\ In the ref. \cite{Weiser} , the main$\mathbf{\ }$goal was to demonstrate
that there is a way to derive the LLG equation coming from a non Hermitian
quantum mechanics and to {\normalsize \ }spark a discussion about the
connection between quantum and classical spin dynamics. Unfortunately, the
quantum Heisenberg equation for a non-Hermitian Hamiltonian operator
describing the damping process is not compatible with the time-evolution
operators for non-Hermitian Hamiltonian operator.

From the relativistic Dirac equation, performing a Foldy-Wouthuysen
transformation and using the Heisenberg equation of spin motion, Mondal et al
\cite{R1,R2,R3} derive general relativistic expressions for the Gilbert
damping, but the term involving the cross-product between the magnetisation
and the time-derivative of the magnetic field is purely imaginary, and
therefore appears not to correspond to damping.

In the seminal work made by Dirac on relativistic quantum mechanics, the
corresponding Hamiltonian would be Hermitian.
We\ stress\ that\ this\ property\ is\ very\ useful\ to\ have\ in\ a\ physical\ system,\ however\ we\ argue\ that\ the\ same\ features\ can\ be\ achieved\ when\ starting\ from\ non-Hermitian\ Hamiltonian\ systems.
These features\ can also be obtained from theories based on non-Hermitian
Hamiltonians that have been considered in different contexts. And\ we
distinguish three separate regimes: i) The\ $\mathcal{PT}$\ -symmetric regime
where the eigenvalues are real, ii) The spontaneously broken $\mathcal{PT}%
$\ regime where the eigenvalues are complex conjugate pairs and iii) The
regime with complex, unrelated, eigenvalues in which the\ $\mathcal{PT}%
$\ -broken regime.\ 

Our objective here is to derive the LLG equation based on a\ non-Hermitian
Dirac Hamiltonian when compared to the most common standard approaches
\cite{R1,R2,R3}.

Therefore, the Dirac equation in its fundamental representation is not unique
to either\textbf{\ }Hermitian quantum mechanics{\normalsize \ }$or$ quantum
field theory. By relaxing the assumption of Hermiticity and adopting instead
the principles of $\mathcal{P}^{0}\mathcal{T}^{0}$-symmetry quantum mechanics
outlined in the following paragraph, we
will\ not\ make\ any\ modifications\ to\textbf{\ }the Dirac equation. By
$\mathcal{P}^{0}\mathcal{T}^{0}$-symmetry we mean reflection in space, with a
simultaneous reversal of time. The fundamental representation of the Dirac
equation emerges completely \textbf{intact,} identical in every aspect to the
Dirac equation derived from Hermitian theory. Before constructing the
analogous 4-d representation using the principles of $\mathcal{P}%
^{0}\mathcal{T}^{0}$\ quantum mechanics, let us briefly recall the notion of
$\mathcal{P}^{0}\mathcal{T}^{0}$-symmetry.

The Hermiticity of quantum Hamiltonians depends on the choice of the inner
product of\ the\ states{\normalsize \ }in the physical Hilbert space. This
point was first pointed out by Bender et al \cite{B1,B2}. They showed that a
wide class of Hamiltonians that respect $\mathcal{P}^{0}\mathcal{T}^{0}%
$-symmetry can exhibit entirely real spectra. Since then $\mathcal{P}%
^{0}\mathcal{T}^{0}$-symmetry has been a subject of intense interest in the
field of quantum mechanics.

While\ any\ evidence\ of\textbf{\ }$\mathcal{P}^{0}\mathcal{T}^{0}$%
\textbf{-}%
symmetry\ has\ remained\ out\ of\ reach\ due\ to\ the\ hermitian\ nature\ of\ the\ quantum\ mechanics\ theory,optics\ have\ provided\ a\ fertile\ ground\ for\ observation\ of\ this\ property-$\mathcal{P}%
^{0}\mathcal{T}^{0}$%
-symmetry-since\ this\ field\ mainly\ relies\ on\ the\ presence\ of\ gain\ and\ loss\textbf{.}%
Note that even though $H$\ and $\mathcal{P}^{0}\mathcal{T}^{0}$\ commute, they
do not continuously have identical eigenvectors, as a result of the
anti-linearity of the $\mathcal{P}^{0}\mathcal{T}^{0}$\ operator. If $H$\ and
$\mathcal{P}^{0}\mathcal{T}^{0}$don't have the same eigenvectors, we say that
the $\mathcal{P}^{0}\mathcal{T}^{0}-$symmetry is broken. The parity operator
$\mathcal{P}^{0}$\ effects the momentum operator $\mathbf{p}$\ and the
position operator $\mathbf{r}$ as ($\mathcal{P}^{0}:\mathbf{r}\rightarrow
-\mathbf{r},$\ $\mathbf{p}\rightarrow-\mathbf{p}$). This parity transformation
has the following effect on the various vector potentials $\mathcal{P}%
^{0}\mathbf{A}(\mathbf{r},t)\left(  \mathcal{P}^{0}\right)  ^{-1}%
=-\mathbf{A}(\mathbf{r},t)$ and $\mathcal{P}^{0}$ $\Phi(\mathbf{r},t)\left(
\mathcal{P}^{0}\right)  ^{-1}=\Phi(\mathbf{r},t)$, expressing thus their
scalar and vector nature.

The anti-linear time reversal operator $\mathcal{T}^{0}$ has the effect of
changing the sign of the momentum operator $\mathbf{p}{\normalsize ,}$ the
pure imaginary complex quantity $i$ and the time $t$\ ($\mathcal{T}%
^{0}:\mathbf{r}\rightarrow\mathbf{r},$\ $\mathbf{p}\rightarrow-\mathbf{p}%
,$\ $i\rightarrow-i,$\ $t\rightarrow-t$). \ Since $\mathbf{A}(\mathbf{r}%
,t)$\ is generated by currents, which reverse\textbf{s\ signs} when the sense
of time is reversed, it holds that $\mathcal{T}^{0}$ $\mathbf{A}%
(\mathbf{r},t)\left(  \mathcal{T}^{0}\right)  ^{-1}=-\mathbf{A}(\mathbf{r},t)$
and $\mathcal{T}^{0}$ $\Phi(\mathbf{r},t)\left(  \mathcal{T}^{0}\right)
^{-1}=\Phi(\mathbf{r},t).$The two reflection operators commute with each
other: $\mathcal{P}^{0}\mathcal{T}^{0}=\mathcal{T}^{0}\mathcal{P}^{0}$.

Therefore, it is natural to introduce a modified Hilbert space, which is now
endowed with $\mathcal{P}^{0}\mathcal{T}^{0}$-inner product, for the
$\mathcal{P}^{0}\mathcal{T}^{0}$-symmetric nonself-adjoint theories. In such a
Hilbert space, the time evolution becomes unitary as the Hamiltonian is
self-$\mathcal{P}^{0}\mathcal{T}^{0}$-adjoint and the eigenfunctions form a
complete set of orthonormal functions. But the norms of the eigenfunctions
have alternate signs even in the new Hilbert space endowed with the
$\mathcal{P}^{0}\mathcal{T}^{0}$-inner products. In fact, any theory having an
unbroken\textbf{\ }$\mathcal{P}^{0}\mathcal{T}^{0}$-symmetry it\ exists a
symmetry of the Hamiltonian associated\ with the fact that there are equal
numbers of positive-norm and negative-norm states \cite{B2}:\textbf{\ }%
\begin{align}
\left\langle \psi_{m},\psi_{n}\right\rangle _{\mathcal{P}^{0}\mathcal{T}^{0}}
&  =\int dx\left[  \mathcal{P}^{0}\mathcal{T}^{0}\psi_{n}(x)\right]  \psi
_{m}(x)=\int dx\psi_{n}^{\ast}(-x)\psi_{m}(x)\nonumber\\
&  =\left(  -1\right)  ^{n}\delta_{mn}\label{ptpt}%
\end{align}

The situation here is analogous to the problem that Dirac encountered in
formulating the spinor wave equation in relativistic quantum theory
\cite{2}\textbf{.}

This again raises an obstacle in probabilistic interpretation in spite of the
system ${\normalsize being}$ in $an${\normalsize \ }unbroken $\mathcal{P}%
^{0}\mathcal{T}^{0}$\ phase. Afterwards, a new symmetry $\mathcal{C}$,
inherent to all $\mathcal{P}^{0}\mathcal{T}^{0}$-symmetric non-Hermitian
Hamiltonians, has been introduced \cite{B2}. $\mathcal{C}$\ commutes with both
$H$\ and $\mathcal{P}^{0}\mathcal{T}^{0}$\ and fixes the problem of negative
norms of the eigenfunctions when the inner products $are${\normalsize \ }taken
with respect to $\mathcal{CP}^{0}\mathcal{T}^{0}$-adjoint.

Does a $\mathcal{P}^{0}\mathcal{T}^{0}$-symmetric Hamiltonian $H$ specify a
physical quantum theory in which the norms of states are positive and time
evolution is unitary? The answer is that if $H$ has an unbroken\textbf{\ }%
$\mathcal{P}^{0}\mathcal{T}^{0}$ symmetry, then it has another symmetry
represented by a linear operator $\mathcal{C}$%
.Therefore\ we\ can\ construct\ a\ time-independent\ inner\ product\ with\ a\ positive-definite\ norm\ in\ terms\ of\textbf{\ }%
$\mathcal{C}$\textbf{.}

Another possibility to explain the reality of the spectrum is making use of
the pseudo/quasi-Hermiticity transformations which do not alter the eigenvalue
spectra.\textrm{\ }It was shown by Mostafazadeh \cite{Mos} that $\mathcal{P}%
^{0}\mathcal{T}^{0}$-symmetric Hamiltonians are only ${\normalsize a}%
$\ specific class of the general families of{\normalsize \ }$the$
pseudo-Hermitian operators. A Hamiltonian is said to be $\eta$%
-pseudo-Hermitian if:%
\begin{equation}
H^{\dagger}=\eta H\eta^{-1},\label{1}%
\end{equation}
where $\eta$ is a \textbf{metric operator}. The eigenvalues of
pseudo-Hermitian Hamiltonians are either real or appear in complex conjugate
pairs ${\normalsize while}$ the eigenfunctions satisfy bi-orthonormality
relations in the conventional Hilbert space. Due to this reason, such
Hamiltonians do not possess ${\normalsize a}$ complete set of orthogonal
eigenfunctions in the conventional Hilbert space and hence the probabilistic
interpretation and unitarity of time evolution have not been satisfied by
these pseudo-Hermitian Hamiltonians.

However, like the case of $\mathcal{P}^{0}\mathcal{T}^{0}$-symmetric
non-Hermitian systems, ${\normalsize the}$ presence of the additional operator
$\eta$\ in the pseudo-Hermitian theories allows ${\normalsize us}$ to define a
new inner product in the fashion
\begin{equation}
\left\langle \phi\right\vert \left.  \psi\right\rangle _{\eta}=\left\langle
\eta\phi\right\vert \left.  \psi\right\rangle =\int\left(  \eta\phi(x)\right)
\psi(x)dx=\left\langle \phi\right\vert \left.  \eta\psi\right\rangle
.\label{IP}%
\end{equation}

Later a novel concept of the pseudoparity-time (pseudo\textbf{-}%
$\mathcal{P}^{0}\mathcal{T}^{0}$\textbf{\ ) }symmetry was introduced in
\cite{mana2020} to connect the non-Hermitian Hamiltonian\textbf{\ }%
$H$\textbf{\ }to its Hermitian conjugate\textbf{\ }$H^{\dagger}$%
\begin{equation}
H^{\dagger}=\left(  \mathcal{P}^{0}\mathcal{T}^{0}\right)  H\left(
\mathcal{P}^{0}\mathcal{T}^{0}\right)  ^{-1}\label{3}%
\end{equation}
where in the expression of the inner product (\ref{IP}), the metric\textbf{\ }%
$\eta$\textbf{\ \ }is replaced by $\mathcal{P}^{0}\mathcal{T}^{0}$. We now
turn our attention to the main topic of interest, the spatial reflection\ and
the time-reversal invariance of the Dirac equation. The complete spatial
reflection (parity) transformation for spinors and the complete time-inversion
operator are denoted $as$ $\mathcal{P}=\gamma^{0}\mathcal{P}^{0}$ and
$\mathcal{T}=\mathcal{-}i\alpha_{1}\alpha_{3}\mathcal{T}^{0}=i\gamma^{1}%
\gamma^{3\text{ }}\mathcal{T}^{0}$ such that \
\begin{align}
\text{ \ }\mathcal{P}\gamma^{0}\mathcal{P}^{-1}  & =\gamma^{0}\text{ ,
}\mathcal{P}\gamma^{i}\mathcal{P}^{-1}=-\gamma^{i}\nonumber\\
\text{ }\mathcal{T}\gamma^{0}\mathcal{T}^{-1}  & =\gamma^{0}\text{ ,
}\mathcal{T}\mathbf{\alpha}\mathcal{T}^{-1}=-\boldsymbol{\alpha}%
\end{align}
where the Hermitian matrices $\ \beta=\gamma^{0}$\ and $\boldsymbol{\alpha}%
$\ satisfy the `Dirac algebra' $\left\{  \alpha_{i},\alpha_{j}\right\}
$\ $=2\delta_{ij}$\ ; $\left\{  \alpha_{i},\beta\right\}  =0$.

In $what${\normalsize \ }$follows,$ we will denote by\
\begin{equation}
\hat{H}^{\mathrm{D}}(t)=i\left(  c\boldsymbol{\alpha}.(\hat{\mathbf{p}%
}+ie\mathbf{A}(\mathbf{r},t))-e\Phi(\mathbf{r},t)+mc^{2}\beta\right)
\label{Hd}%
\end{equation}
the non-Hermitian Dirac Hamiltonian for a single electron in
${\normalsize the}$ presence of a classical time-dependent external
electromagnetic field defined by $\left(  \mathbf{A}(\mathbf{r},t),\Phi
(\mathbf{r},t)\right)  $. \ The associated non-Hermitian Dirac equation for a
single electron in ${\normalsize the}$\ presence of a classical time-dependent
external electromagnetic field reads
\begin{align}
i\hbar\frac{\partial\Psi^{\mathrm{D}}(\mathbf{r},t)}{\partial t}  & =\hat
{H}^{\mathrm{D}}(t)\Psi^{\mathrm{D}}(\mathbf{r},t)\nonumber\\
& =i\left(  \hat{T}+\hat{V}+mc^{2}\beta\right)  \Psi^{\mathrm{D}}%
(\mathbf{r},t)\;,\label{eq1}%
\end{align}
where $\Psi^{\mathrm{D}}(r,t)=\operatorname{col}(u(r,t),v(r,t)),$\ is
bispinors\ verifying the pseudo-othogonality relation $\left\langle
\Psi^{\mathrm{D}}|\Psi^{\mathrm{D}}\right\rangle _{\mathcal{PT}}=I$, $\hat
{V}\equiv-e\Phi$ and $\hat{T}\equiv c\mathbf{\alpha}.\hat{\mathbf{\pi}%
}=c\mathbf{\alpha}.(\hat{\mathbf{p}}+ie\mathbf{A}(\mathbf{r},t))$ is the
kinetic energy which \textbf{produces} a coupling between small and large
components of the Dirac wavefunction $\Psi^{\mathrm{D}},$ $\mathbf{\alpha}$,
$\beta$ and $\gamma^{5}$ (see
Eq.(\ref{Heisenberg equation of motion for the spin mean-value})) are the
Dirac matrices \cite{Str05}. The $\mathcal{PT}$ symmetry condition
\begin{equation}
\left(  \mathcal{PT}\right)  \hat{H}^{\mathrm{D}}(t)\left(  \mathcal{PT}%
\right)  ^{-1}=\hat{H}^{\dag\mathrm{D}}(t),\label{pse-PT}%
\end{equation}
connects the non-Hermitian Dirac Hamiltonian $\hat{H}^{\mathrm{D}}(t)$ to its
Hermitian conjugate $\hat{H}^{\dag\mathrm{D}}(t).$This observation leads us to
introduce a novel concept of the pseudo-parity-time (pseudo-$\mathcal{PT}$ )
symmetry, where $\left(  \mathcal{PT}\right)  $ is interpreted as a metric.
Thus as the case of pseudo-hermiticity, the bispinor\ $\Psi^{\mathrm{D}%
}(r,t)=\operatorname{col}(u(r,t),v(r,t)),$\ \ verif$ies$ the
pseudo-othogonality relation $\left\langle \Psi^{\mathrm{D}}|\Psi^{\mathrm{D}%
}\right\rangle _{\mathcal{PT}}=I$ .\textrm{\ }Note that, there is another
situation which differs from that described above\textbf{,} also called
pseudo-$\mathcal{PT}$ \ symmetry which means that the system can have a real
eigenvalues whether or not the original system is $\mathcal{PT}$- symmetric
\cite{9,10}.

As we are dealing with the non-relativistic expansion of the Dirac equation,
the following decomposition $\Psi^{\mathrm{D}}(\mathbf{r},t)=\Pi
(\mathbf{r},t)\times\chi^{\mathrm{D}}(\mathbf{u}_{s},t)$ \cite{Rom04} can be
used, where $\chi^{\mathrm{D}}(\mathbf{u}_{s},t)$ is the time-dependent
bi-spinor representing the spin state oriented in the direction defined by
$\mathbf{u}_{s}$ and $\Pi(\mathbf{r},t)$ the \textbf{scalar} part of the wave function.

We argue that, in the non-relativistic limit, the operator $\hat
{\overline{\mathbf{\Sigma}}}^{\mathrm{D}}$\ is the one which must be
interpreted as the spin operator in the Pauli theory and used to define the
magnetization as $\mathbf{M}(\mathbf{r},t)\equiv\mu_{\mathrm{B}}\left\langle
\chi^{\mathrm{D}}|\hat{\overline{\mathbf{\Sigma}}}^{\mathrm{D}}|\chi
^{\mathrm{D}}\right\rangle _{\mathcal{PT}}=\mu_{\mathrm{B}}\left\langle
\chi^{\mathrm{D}}|\hat{U}^{-1}\hat{\overline{\mathbf{\Sigma}}}^{\mathrm{FW}%
}\hat{U}|\chi^{\mathrm{D}}\right\rangle _{\mathcal{PT}}=\mu_{\mathrm{B}%
}\langle\chi^{\mathrm{FW}}|\mathbf{\hat{\Sigma}}^{\mathrm{FW}}|\chi
^{\mathrm{FW}}\rangle_{\mathcal{PT}}$ where $\chi^{\mathrm{FW}}$ is the spin
part of the{\normalsize \ }Dirac{\normalsize \ }bi-spinor{\normalsize \ }%
wave{\normalsize \ }function{\normalsize \ }$\Psi^{\mathrm{D}}$%
{\normalsize \ i}n a{\normalsize \ }non-relativistic{\normalsize \ }%
\textit{expansion}${\normalsize ,}$ obtained by using the Foldy-Wouthuysen
(FW) transformation \cite{Fol50} and $\hat{U}$\ is the associated operator
(see the definition in the following). It worth mentioning
that${\normalsize ,}$ according to the above definition${\normalsize ,}$
expanding the spin operator (to a consistent order in $h/mc$) and using the
Dirac representation of the wave function is equivalent to
expand${\normalsize ing}$ the wave function (also to a consistent order in
$h/mc$ using the FW transformation) and keep$ing${\normalsize \ }the original
spin operator in the Dirac representation. In this work we have chosen to
expand the mean value operator. The classical magnetization, $\mathbf{M}%
(\mathbf{r},t)$, is obtained by using the correspondence principle. Indeed, we
will show in what follows that the equation of motion of the mean spin
operator for an electron interacting with a time-dependent electromagnetic
field leads to the LLG equation of motion revealing thus its microscopic origin.

In a seminal work, Foldy and Wouthuysen (FW) solved the problem of finding a
canonical transformation $that${\normalsize \ }$allows$ to obtain a
two-component theory in the low-energy limit (Pauli
approximation)${\normalsize ,}${\normalsize \ }in the case of the Dirac
equation coupled to an electromagnetic field \cite{Fol50}. Unfortunately,
contrary to the free-electron case, the solution cannot be expressed in a
closed form. However, FW showed how to obtain successive approximations of
this transformation as a power series expansion in powers of the Compton wave
length of the particle $\lambda_{\mathrm{C}}\equiv\frac{h}{mc}$. This
procedure, generally restricted to the second-order in $1/m$, is presented in
many textbook${\normalsize s}$ on relativistic quantum mechanics
\cite{Str05,Itz85,Gre00,Rei09} and has been extended to fifth order in powers
of $1/m$ \cite{Her12}.

In what follows, the symbol $[\hat{C},\hat{D}]$ ($\{\hat{C},\hat{D}\}$)
denotes the commutator (anticommutator) of the operators $\hat{C}$ and
$\hat{D}$. We shall also use the following notations: $\hat{X}\equiv
X(\mathbf{r},t)$ and $\hat{Y}\equiv Y(\mathbf{r},t)$.

In the Hermitian Dirac representation
\begin{equation}
\hat{H}^{h\mathrm{D}}(t)=\left(  c\boldsymbol{\alpha}.(\hat{\mathbf{p}%
}-e\mathbf{A}(\mathbf{r},t))+e\Phi(\mathbf{r},t)+mc^{2}\beta\right)  ,
\end{equation}
the Heisenberg equation of motion for the spin operator reads as follows
\begin{align}
\frac{d\hat{\mathbf{\Sigma}}^{\mathrm{D}}}{dt}  & =\frac{i}{\hbar}\left[
\hat{H}^{h\mathrm{D}},\hat{\mathbf{\Sigma}}^{\mathrm{D}}\right] \nonumber\\
& =-2\frac{c}{\hbar}\left[  \mathbf{\alpha}\wedge(\hat{\mathbf{p}}%
-e\mathbf{A}(\mathbf{r},t)\right]  \;.\label{eq2}%
\end{align}

It is well established that the expectation value onto $\left\vert
\Psi^{\mathrm{D}}\right\rangle $\ of the above equation does not lead to the
LLG equation for the magnetization. However, as it will ${\normalsize be}$
shown in the following, the latter can be obtained by using the non unitary FW
transformation and the new definition of the magnetization as an expectation
value of mean spin operator. The first- and second-order terms of the FW
expansion in powers of $1/m$ correspond, respectively, to the precessional
motion of the magnetization around an effective magnetic field and its damping.

Since the Hamiltonian\textbf{\ }$\hat{H}^{\mathrm{D}}(t)$\textbf{\ }has a
similar structure to\ the\ one in the Dirac case, by analogy with the latter,
we use for\textbf{\ }$\hat{U}(t)$\textbf{\ }the form\textbf{\ }$e^{\hat{S}%
(t)}$ \textbf{\ }where\textbf{\ }$\hat{S}$ \textbf{\ }is a non self-adjoint
operator. Therefore, the transformation $\Psi^{\mathrm{FW}}(r,t)=e^{\hat
{S}(t)}\Psi^{\mathrm{D}}(r,t)\equiv\hat{U}(t)\Psi^{\mathrm{D}}(r,t)$ leads to
a new Hamiltonian
\begin{equation}
\hat{H}^{\mathrm{FW}}(t)=e^{\hat{S}(t)}\left(  \hat{H}^{\mathrm{D}}%
(t)-i\hbar\frac{\partial}{\partial t}\right)  e^{-\hat{S}(t)}\;,\label{eq2'}%
\end{equation}
where $\hat{S}$ is a non self-adjoint operator. More generally, any operator
\textit{in} the FW representation${\normalsize ,}$ that is not explicitly time
dependent, $\hat{O}^{\mathrm{FW}}$ will be transformed in the Dirac
representation as $\hat{O}^{\mathrm{D}}(t)=\hat{U}^{-1}(t)\hat{O}%
^{\mathrm{FW}}\hat{U}(t)$.\ 

\ The most natural extension of the Ehrenfest equation to non-Hermitian
pseudo-$\mathcal{PT}$ symmetric systems is $by${\normalsize \ }$replacing$ a
Hermitian $\hat{H}^{h\mathrm{D}}$\ ${\normalsize with}\mathcal{\ }$ a
non-Hermitian one. The structure of the Ehrenfest\ equation does not change,
having assumed that part of the action of $\mathcal{T}$\ (i.e. $\mathcal{T}$
$^{0}$) is to send $t\rightarrow-t$, i.e.; $we$ $show$ that it
anticommute${\normalsize s}$ with the operator $\partial/\partial t$,
consequently, the operator $\mathcal{PT}$ commute${\normalsize s}$ with
$i\partial/\partial t$ . $Which${\normalsize \ }$immediately${\normalsize \ }%
$leads\ us${\normalsize \ }$to$ deduce the Ehrenfest equation of motion for
the diagonal matrix element of an operator $\hat{O}^{\mathrm{D}}(t)$%
\begin{equation}
\frac{d}{dt}\left\langle \Psi^{\mathrm{D}}|\hat{O}^{\mathrm{D}}|\Psi
^{\mathrm{D}}\right\rangle _{\mathcal{PT}}=\left\langle \Psi^{\mathrm{D}%
}|\frac{i}{\hbar}\left[  \hat{H}^{\mathrm{D}},\hat{O}^{\mathrm{D}}\right]
+\frac{\partial\hat{O}^{\mathrm{D}}}{\partial t}|\Psi^{\mathrm{D}%
}\right\rangle _{\mathcal{PT}}.\label{eq2"}%
\end{equation}
It's straightforward to show that the equation of motion \ (\ref{eq2"}) leads
to%
\begin{equation}
\frac{d}{dt}\left\langle \Psi^{\mathrm{FW}}|\hat{O}^{\mathrm{FW}}%
|\Psi^{\mathrm{FW}}\right\rangle _{\mathcal{PT}}=\left\langle \Psi
^{\mathrm{FW}}|\frac{i}{\hbar}\left[  \hat{H}^{\mathrm{FW}},\hat
{O}^{\mathrm{FW}}\right]  |\Psi^{\mathrm{FW}}\right\rangle _{\mathcal{PT}%
}.\label{eq2'''}%
\end{equation}
By expanding $\hat{S}=\hat{S}_{1}m^{-1}+\hat{S}_{2}m^{-2}+\hat{S}_{3}%
m^{-3}+...$ with $\hat{S}_{1}=\frac{\beta}{2c}\left(  \mathbf{\alpha}%
.\hat{\mathbf{\pi}}\right)  $, $\hat{S}_{2}=-\frac{i\hbar e}{4c^{3}}\left(
\boldsymbol{\alpha}.\mathbf{E}\right)  $, $\hat{S}_{3}=-\frac{\beta}{8c^{6}%
}\left(  \frac{4c^{3}}{3}\left(  \mathbf{\alpha}.\hat{\mathbf{\pi}}\right)
\left(  \mathbf{\alpha}.\hat{\mathbf{\pi}}\right)  \left(  \mathbf{\alpha
}.\hat{\mathbf{\pi}}\right)  +iec\hbar^{2}\left(  \mathbf{\alpha}.\partial
_{t}\mathbf{E}\right)  \right)  $ and $\mathbf{E}=-\nabla\Phi-\frac
{\partial\mathbf{A}}{\partial t}$ \cite{Her12}, the mean spin operator is
computed using the inverse FW transformation of the spin operator as
$\hat{\overline{\mathbf{\Sigma}}}^{\mathrm{D}}=e^{-\left(  \hat{S}_{1}%
/m+\hat{S}_{2}/m^{2}+\hat{S}_{3}/m^{3}\right)  }\hat{\mathbf{\Sigma}%
}^{\mathrm{D}}e^{\left(  \hat{S}_{1}/m+\hat{S}_{2}/m^{2}+\hat{S}_{3}%
/m^{3}\right)  }$ and may be expanded in power series of $(1/m)$ leading to
$\hat{\overline{\mathbf{\Sigma}}}^{\mathrm{D}}=\hat{\overline{\mathbf{\Sigma}%
}}_{0}^{\mathrm{D}}+\hat{\overline{\mathbf{\Sigma}}}_{1}^{\mathrm{D}}%
m^{-1}+\hat{\overline{\mathbf{\Sigma}}}_{2}^{\mathrm{D}}m^{-2}+\hat
{\overline{\mathbf{\Sigma}}}_{3}^{\mathrm{D}}m^{-3}+...$%

\begin{align}
\hat{\overline{\mathbf{\Sigma}}}_{0}^{\mathrm{D}}  & =\hat{\mathbf{\Sigma}%
}^{\mathrm{D}}\;,\nonumber\\
\hat{\overline{\mathbf{\Sigma}}}_{1}^{\mathrm{D}}  & =-\frac{i\beta}{c}\left(
\mathbf{\alpha}\times\hat{\mathbf{\pi}}\right)  \;,\nonumber\\
\hat{\overline{\mathbf{\Sigma}}}_{2}^{\mathrm{D}}  & =\frac{1}{8c^{2}}\left(
-4ie\hbar\mathbf{B}+2e\hbar\left(  \hat{\mathbf{\Sigma}}^{\mathrm{D}}%
\times\mathbf{B}\right)  -4\left(  \hat{\mathbf{\pi}}\times\left(
\hat{\mathbf{\Sigma}}^{\mathrm{D}}\times\hat{\mathbf{\pi}}\right)  \right)
\right)  -\frac{e\hbar}{2c^{3}}\left(  \boldsymbol{\alpha}\times
\mathbf{E}\right)  \;,\nonumber\\
\hat{\overline{\mathbf{\Sigma}}}_{3}^{\mathrm{D}}  & =\frac{\beta}{48c^{3}%
}\left[  \left(  \mathbf{\alpha}.\hat{\mathbf{\pi}}\right)  ,\left[  \left(
\mathbf{\alpha}.\hat{\mathbf{\pi}}\right)  ,\left[  \left(  \mathbf{\alpha
}.\hat{\mathbf{\pi}}\right)  ,\hat{\mathbf{\Sigma}}^{\mathrm{D}}\right]
\right]  \right]  +\frac{\beta}{6c^{3}}\left[  \left(  \mathbf{\alpha}%
.\hat{\mathbf{\pi}}\right)  \left(  \mathbf{\alpha}.\hat{\mathbf{\pi}}\right)
\left(  \mathbf{\alpha}.\hat{\mathbf{\pi}}\right)  ,\hat{\mathbf{\Sigma}%
}^{\mathrm{D}}\right] \nonumber\\
& -\frac{e\hbar^{2}\beta}{4c^{5}}\left(  \mathbf{\alpha\times}\left(
\partial_{t}\mathbf{E}\right)  \right)  ,\label{spin-mv-third}%
\end{align}
where $\mathbf{B}=\nabla\times\mathbf{A}.$ The free case which is investigated
in \cite{Fol50} (may be obtained in closed form in this case) is recovered
from the above formula by substituting $\mathbf{E}=\mathbf{B}=0$ ,
$i\beta\rightarrow\beta$ and $i\hat{\mathbf{\pi}}\rightarrow\hat{\mathbf{p}}$
leading to
\[
\hat{\overline{\mathbf{\Sigma}}}^{\mathrm{D}}=\hat{\mathbf{\Sigma}%
}^{\mathrm{D}}-\frac{i\beta\left(  \mathbf{\alpha}\wedge\hat{\mathbf{p}%
}\right)  }{E_{p}}-\frac{\left(  \hat{\mathbf{p}}\wedge\left(  \hat
{\mathbf{\Sigma}}^{\mathrm{D}}\wedge\hat{\mathbf{p}}\right)  \right)  }%
{E_{p}\left(  E_{p}+mc\right)  }%
\]
where $E_{p}=\sqrt{m^{2}c^{2}+p^{2}}$.

The pseudo-$\mathcal{PT}$ equation of motion (\ref{eq2"}) for the mean spin
operator is%

\begin{align}
& \frac{d}{dt}\left\langle \Psi^{\mathrm{D}}\right\vert \hat{\overline
{\mathbf{\Sigma}}}^{\mathrm{D}}\left\vert \Psi^{\mathrm{D}}\right\rangle
_{\mathcal{PT}}\nonumber\\
& =\left\langle \Psi^{\mathrm{D}}\right\vert \frac{e\beta}{m}\left(
\hat{\mathbf{\Sigma}}^{\mathrm{D}}\wedge\mathbf{B}\right)  -\frac{e\hbar
}{4m^{2}c^{2}}\left(  \hat{\mathbf{\Sigma}}^{\mathrm{D}}\wedge\partial
_{t}\mathbf{B}\right)  -\frac{ie}{2m^{2}c^{2}}\left(  \hat{\mathbf{\Sigma}%
}^{\mathrm{D}}\wedge\left(  \mathbf{E}\wedge\hat{\mathbf{\pi}}\right)  \right)
\nonumber\\
& +\frac{1}{4\hbar m^{2}c}\left(  \left[  \left(  \mathbf{\alpha}.\mathbf{\pi
}\right)  \hat{\Sigma}^{\mathrm{D}}\left(  \mathbf{\alpha}.\mathbf{\pi
}\right)  ,\left(  \mathbf{\alpha}.\mathbf{\pi}\right)  \right]  -\left[
\left(  \mathbf{\alpha}.\mathbf{\pi}\right)  \left(  \mathbf{\alpha
}.\mathbf{\pi}\right)  \left(  \mathbf{\alpha}.\mathbf{\pi}\right)
,\Sigma^{\mathrm{D}}\right]  \right)  +\vartheta(m^{-3})\left\vert
\Psi^{\mathrm{D}}\right\rangle _{\mathcal{PT}}%
.\label{Heisenberg equation of motion for the spin mean-value}%
\end{align}
$\mathcal{(}$In order$\mathcal{)}$ to check this result, we have applied to
the above equation the direct FW transformation. It leads to%
\begin{align}
\frac{d}{dt}\left\langle \Psi^{\mathrm{D}}\right\vert \hat{\overline
{\mathbf{\Sigma}}}^{\mathrm{D}}\left\vert \Psi^{\mathrm{D}}\right\rangle
_{\mathcal{PT}}  & =\frac{d}{dt}\left\langle \Psi^{\mathrm{FW}}\right\vert
e^{\left(  \hat{S}_{1}/m+\hat{S}_{2}/m^{2}+\hat{S}_{3}/m^{3}\right)  }%
\hat{\overline{\mathbf{\Sigma}}}^{\mathrm{D}}e^{-\left(  \hat{S}_{1}/m+\hat
{S}_{2}/m^{2}+\hat{S}_{3}/m^{3}\right)  }\left\vert \Psi^{\mathrm{FW}%
}\right\rangle _{\mathcal{PT}}\nonumber\\
& \equiv\frac{d}{dt}\left\langle \Psi^{\mathrm{FW}}\right\vert \mathbf{\hat
{\Sigma}}\left\vert \Psi^{\mathrm{FW}}\right\rangle _{\mathcal{PT}%
}=\left\langle \Psi^{\mathrm{FW}}\right\vert \frac{i}{\hbar}\left[  \hat
{H}^{\mathrm{FW}},\hat{\overline{\mathbf{\Sigma}}}^{\mathrm{FW}}\right]
\left\vert \Psi^{\mathrm{D}}\right\rangle _{\mathcal{PT}}\label{Sp}%
\end{align}
where $\mathbf{\hat{\Sigma}}^{\mathrm{FW}}\equiv$ $\mathbf{\hat{\Sigma}}$ and
the expression of $\hat{H}^{\mathrm{FW}}$ is given by\
\begin{align}
\hat{H}^{\mathrm{FW}}  & =i\beta mc^{2}-ie\hat{\Phi}+i\beta\frac
{\hat{\mathbf{\pi}}^{2}}{2m}+i\beta\frac{\hat{\mathbf{\pi}}^{4}}{8m^{3}c^{2}%
}+\beta\frac{e\hbar^{2}}{16m^{3}c^{4}}\{\hat{\mathbf{\pi}},\partial
_{t}\mathbf{E}\}\nonumber\\
& -\beta\frac{e\hbar}{2m}\hat{\mathbf{\Sigma}}.\left[  \mathbf{B}-\frac
{\beta\hbar}{4mc^{2}}\left(  \left(  \mathbf{\nabla\wedge E}\right)
+\frac{2i}{\hbar}\mathbf{E}\wedge\hat{\mathbf{\pi}}\right)  +\frac{i\hbar
}{8m^{2}c^{4}}\left(  \partial_{t}\mathbf{E}\wedge\hat{\mathbf{\pi}}%
)+\hat{\mathbf{\pi}}\wedge\partial_{t}\mathbf{E}\right)  \right] \nonumber\\
& -\beta\frac{e\hbar}{8m^{3}c^{2}}\left\{  \hat{\mathbf{\pi}}^{2}%
,\hat{\mathbf{\Sigma}}.\mathbf{B}\right\}  -i\beta\left(  \frac{e\hbar}%
{2m}\right)  ^{2}\frac{\mathbf{B}^{2}}{2mc^{2}}+\frac{ie\hbar^{2}}{8m^{2}%
c^{2}}\mathbf{\nabla}.\mathbf{E}+\vartheta(m^{-4}).\label{HFW}%
\end{align}

Now, the equation of motion for the mean spin operator
(\ref{Heisenberg equation of motion for the spin mean-value}) can be written
in a simpler way being obtained as the pseudo-$\mathcal{PT}$ expectation value
a spin Dirac states $|\chi^{\mathrm{D}}\rangle$ associated to $\hat
{H}^{\mathrm{D}}$, one gets
\begin{align}
& \frac{d}{dt}\langle\chi^{\mathrm{D}}|\hat{\overline{\mathbf{\Sigma}}%
}^{\mathrm{D}}|\chi^{\mathrm{D}}\rangle_{\mathcal{PT}}\nonumber\\
& =\frac{e\beta}{m}\langle\chi^{\mathrm{D}}|\hat{\mathbf{\Sigma}}^{\mathrm{D}%
}|\chi^{\mathrm{D}}\rangle_{\mathcal{PT}}\wedge\mathbf{B}-\frac{e\hbar}%
{4m^{2}c^{2}}\langle\chi^{\mathrm{D}}|\hat{\mathbf{\Sigma}}^{\mathrm{D}}%
|\chi^{\mathrm{D}}\rangle_{\mathcal{PT}}\wedge\partial_{t}\mathbf{B}%
\nonumber\\
& -\frac{e}{2mc^{2}}\langle\chi^{\mathrm{D}}|\hat{\mathbf{\Sigma}}%
^{\mathrm{D}}|\chi^{\mathrm{D}}\rangle_{\mathcal{PT}}\wedge\left(
\mathbf{E}\wedge\langle\chi^{\mathrm{D}}|\frac{i}{m}\hat{\mathbf{\pi}}%
|\chi^{\mathrm{D}}\rangle_{\mathcal{PT}}\right)  +\vartheta(m^{-3}%
)\;.\;\label{eq_meanvalue1}%
\end{align}

The above expression has been obtained by using $\left\langle \Psi
^{\mathrm{D}}\right\vert \hat{\overline{\mathbf{\Sigma}}}^{\mathrm{D}%
}\left\vert \Psi^{\mathrm{D}}\right\rangle _{\mathcal{PT}}=\langle
\chi^{\mathrm{D}}|\hat{\overline{\mathbf{\Sigma}}}^{\mathrm{D}}|\chi
^{\mathrm{D}}\rangle_{\mathcal{PT}}$. It is of interest to mention that the
non-diagonal terms which appear at the \textbf{second} line of
(\ref{Heisenberg equation of motion for the spin mean-value}) are due to the
Zitterbewegung phenomenon \cite{Sak67}. They cancel out when they are pseudo
averaged out in a Dirac states.

Let us note\ that, it is more convenient to use the FW representation to find
the evolution of spin, $\mathcal{(}$indeed$\mathcal{)}$ from Eq.(\ref{eq1})
where $\frac{d}{dt}\left\langle \Psi^{\mathrm{D}}\right\vert \hat
{\overline{\mathbf{\Sigma}}}^{\mathrm{D}}\left\vert \Psi^{\mathrm{D}%
}\right\rangle _{\mathcal{PT}}\equiv\frac{d}{dt}\left\langle \Psi
^{\mathrm{FW}}\right\vert \mathbf{\hat{\Sigma}}^{\mathrm{FW}}\left\vert
\Psi^{\mathrm{FW}}\right\rangle _{\mathcal{PT}}$, we get%
\begin{equation}
\frac{d}{dt}\langle\chi^{\mathrm{FW}}|\mathbf{\hat{\Sigma}}|\chi^{\mathrm{FW}%
}\rangle_{\mathcal{PT}}=\frac{e\beta}{m}\langle\chi^{\mathrm{FW}}%
|\mathbf{\hat{\Sigma}}\wedge\left(  \mathbf{B}+\frac{1}{2c^{2}}\left[
\mathbf{E}\wedge\frac{i\beta}{m}\hat{\mathbf{\pi}}\right]  \right)
-\frac{e\hbar}{4m^{2}c^{2}}\left[  \mathbf{\hat{\Sigma}}\wedge\left(
\mathbf{\nabla\wedge E}\right)  \right]  |\chi^{\mathrm{FW}}\rangle
_{\mathcal{PT}}\label{24}%
\end{equation}

Concerning the damping process, as previously explained, the only term of
importance in the expression
(\ref{Heisenberg equation of motion for the spin mean-value}) is
$-\frac{e\hbar}{4m^{2}c^{2}}\left(  \mathbf{\hat{\Sigma}}^{\mathrm{D}}%
\wedge\partial_{t}\mathbf{B}\right)  $ which has been obtained from the
commutator $\left[  \left(  \mathbf{\alpha}\times\mathbf{E}\right)  ,\left(
\mathbf{\alpha}.\hat{\mathbf{\pi}}\right)  \right]  $ coming from $\left[
\hat{H}^{\mathrm{D}},\mathbf{\hat{\Sigma}}^{\mathrm{D}}\right]  $ and the
partial derivative with respect to time of the mean spin angular momentum
operator at second order in $1/m$ \cite{Note1} given in Eq.
(\ref{spin-mv-third}). In addition, classical Maxwell equations have been also
employed. Similarly to the Breit Hamiltonian$\mathcal{,}$ which is obtained
from the classical Darwin Lagrangian (which also originates from Maxwell
equations) by using the correspondence principle (CP) \cite{Rei09}%
${\normalsize ,}$ we here resort to the same procedure (in its inverse form,
from quantum to classical) for the Maxwell equations. According to this
principle, the quantum counterparts $\hat{f}$, $\hat{g}$ of classical
observables $f$, $g$ satisfy $\left\langle \left[  \mathrm{\ }\hat{f},\hat
{g}\right]  \right\rangle =i\hbar\left\{  f,g\right\}  _{p,q}$
where\textrm{\ }$\left\langle \left[  \mathrm{\ }\hat{f},\hat{g}\right]
\right\rangle $\textrm{\ }is the expectation value of the commutator and the
symbol $\left\{  {}\right\}  _{p,q}$ denotes the Poisson bracket
\cite{Lib87,Jean2,Hove,D. Sen}. Let's take for instance the Maxwell-Faraday
equation, we have%
\begin{equation}
\mathbf{\nabla}\wedge\mathbf{E}(\mathbf{r},t)=-\frac{\partial\mathbf{B}%
(r,t)}{\partial t}%
\end{equation}
which can be rewritten as
\begin{equation}
\epsilon_{ijk}\left\{  p_{i},E_{j}\right\}  _{p,q}e_{k}=-\frac{\partial
\mathbf{B}(\mathbf{r},t)}{\partial t}\;,
\end{equation}
and using the CP one gets
\begin{equation}
\epsilon_{ijk}\frac{\left[  \hat{p}_{i},E_{j}\right]  }{i\hbar}e_{k}%
=-\frac{\partial\mathbf{B}(\mathbf{r},t)}{\partial t}\equiv-\partial
_{t}\mathbf{B}\;.
\end{equation}

Consequently, by using our definition of the \textbf{magnetization}
$\mathbf{M}(\mathbf{r},t)\equiv\mu_{\mathrm{B}}\left\langle \chi^{\mathrm{D}%
}|\hat{\overline{\mathbf{\Sigma}}}^{\mathrm{D}}|\chi^{\mathrm{D}}\right\rangle
_{\mathcal{PT}}\equiv\mu_{\mathrm{B}}\left\langle \chi^{\mathrm{FW}%
}|\mathbf{\hat{\Sigma}}|\chi^{\mathrm{FW}}\right\rangle _{\mathcal{PT}}%
$\ \ and the CP, the equation of motion (\ref{eq_meanvalue1}) may be rewritten
for the electron part as
\begin{align}
\frac{d\mathbf{M}(\mathbf{r},t)}{dt}  & =\frac{e}{m}\mathbf{M}(\mathbf{r}%
,t)\wedge\mathbf{B}(\mathbf{r},t)-\frac{e}{4m^{2}c^{2}}\mathbf{M}%
(\mathbf{r},t)\wedge\mathcal{\partial}_{t}\mathbf{B}(\mathbf{r},t)\nonumber\\
& +\frac{e}{2mc^{2}}\mathbf{M}(\mathbf{r},t)\wedge\left(  \mathbf{E}%
(\mathbf{r},t)\wedge\mathbf{v}\right)  +\vartheta(m^{-3}).\label{eq_meanvalue}%
\end{align}
The above equation constitutes the main result of this work.

Moreover, if the electron is embedded in a magnetically polarizable
medium${\normalsize ,}$ defined by its magnetic polarizability $\chi
_{m}{\normalsize ,}$ then $\frac{\partial\mathbf{M}}{\partial t}$ generates a
time-dependent magnetic induction according to the relation $\partial
_{t}\mathbf{B}(\mathbf{r},t)=\frac{1}{\chi_{m}}\frac{\partial\mathbf{M}%
}{\partial t}$ and the equation (\ref{eq_meanvalue}) can be rewritten as
\begin{equation}
\frac{d\mathbf{M}(\mathbf{r},t)}{dt}=-\gamma\mathbf{M}(\mathbf{r}%
,t)\wedge\mathbf{B}_{\mathrm{eff}}(\mathbf{r},t)-\frac{\alpha_{\mathrm{G}}}%
{M}\left(  \mathbf{M}(\mathbf{r},t)\wedge\frac{\partial\mathbf{M}%
(\mathbf{r},t)}{\partial t}\right) \label{LLG}%
\end{equation}
with $\gamma=-\frac{e}{m}>0$ the gyromagnetic ratio for an isolated electron,
$\mathbf{B}_{\mathrm{eff}}\equiv\mathbf{B}-\frac{1}{2c^{2}}\mathbf{v}%
\wedge\mathbf{E}$ and $\alpha_{\mathrm{G}}\equiv\frac{eM}{4m^{2}c^{2}\chi_{m}%
}$. The first term describes the precessional motion of the magnetization
vector around the direction of the effective magnetic field and the second
term represents its damping$,$ characterized by the Gilbert's constant
$\alpha_{\mathrm{G}}$.

Let us stress that the first term in the right hand side of equation
(\ref{LLG}) can be retrieved from the non-relativistic expansion of the
Bargmann-Michel-Telegdi's equation \cite{Itz85,BMT59,Jac98} which
\textbf{represents} the relativistic equation of motion of a classical
magnetic dipole moment \cite{Note2}. However, the damping term cannot be
obtained from this classical description due to its quantum origin.

In summary, the mean spin angular momentum operator introduced for the first
time by Foldy and Wouthuysen for the case of a free electron has been extended
to the non -Hermitian or precisely to a pseudo\textbf{\ }$\mathcal{PT}%
$\textbf{-}symmetric case of an electron interacting with a time-dependent
electromagnetic field. The \textbf{expectation} equation of the motion of the
latter leads to the Landau-Lifshitz-Gilbert equation revealing thus its
microscopic origin. We therefore argue that the expectation value of the
pseudo-mean spin operator with the new definition of\textbf{\ }$\mathcal{PT}%
$\textbf{-}inner product must be used instead of the usual one to properly
describe the dynamics of the spin magnetization.

\textbf{Supplementary Materials}

In terms of the conjugate variable $(q,p)$ the classical spin $\overrightarrow
{S}$ is desribed by \cite{S1,S2}%
\begin{equation}
\left\{
\begin{array}
[c]{c}%
S_{x}=\sqrt{S^{2}-p^{2}}\cos q\\
S_{y}=\sqrt{S^{2}-p^{2}}\sin q\\
S_{z}=p
\end{array}
\right. \label{sp}%
\end{equation}
the Poisson brackets $\left\{  S_{i},S_{j}\right\}  =\varepsilon
_{ijk}S_{k\text{ }}$($i,j,k$ are $x,y$ or $z$) are analogous to the same
relationships one has with spin components and commutators in quantum mechanics.

\bigskip Suppose we have the following Hamiltonian
\begin{equation}
H=\overrightarrow{B}.\overrightarrow{S}\label{h}%
\end{equation}
which is formally identical to the Hamiltonian for a spin $1/2$ system in a
uniform magnetic field. We can calculate the evolution of the vector
components using the standard Hamiltonian techniques and The motion of spin
$\overrightarrow{S}$ on the sphere (phase space) with (conserved) radius $S$=
$\left\vert \overrightarrow{S}\right\vert $ generated by (\ref{h}), can be
obtained by regarding $H$ (\ref{h}) as classical hamiltonian . It may be
confirmed that Hamilton's equation reproduce exactly what spin does in a
magnetic field i.e, $\overrightarrow{\overset{\cdot}{S}}=\overrightarrow
{B}\wedge\overrightarrow{S}$.

The two-level spin system can be written as a classical model if we employ the
anticommuting Grassmann variables \cite{S3,S4,S5,S6} $\overrightarrow{\zeta}$
which are transformed to the spin operator after the quantization
$\overset{\wedge}{\overrightarrow{\zeta}}=$ $\overset{\wedge}{\overrightarrow
{S}}/\sqrt{2}$ . Unlike the classical spin defined in the equation
((\ref{sp})) which does not tranformed into a spin operator after the
quantization $\overset{\wedge}{\overrightarrow{S}}\neq\overrightarrow{S}.$

\end{document}